\documentclass[journal=jctcce,manuscript=article,layout=twocolumn]{achemso}

\usepackage[english]{babel}
\usepackage[table]{xcolor}

\usepackage{amsmath}
\usepackage{graphicx}
\usepackage{natbib}

\usepackage[colorlinks=true, allcolors=blue]{hyperref}

\title{Enhancing Protein-Ligand Binding Affinity Predictions using Neural Network Potentials}

\author{Francesc Saban\'es Zariquiey}
\affiliation{Computational Science Laboratory, Universitat Pompeu Fabra, Barcelona Biomedical Research Park (PRBB), C Dr. Aiguader 88, 08003, Barcelona, Spain}
\alsoaffiliation{Acellera Labs, C Dr Trueta 183, 08005, Barcelona, Spain}
\author{Raimondas Galvelis}
\affiliation{Computational Science Laboratory, Universitat Pompeu Fabra, Barcelona Biomedical Research Park (PRBB), C Dr. Aiguader 88, 08003, Barcelona, Spain}
\alsoaffiliation{Acellera Labs, C Dr Trueta 183, 08005, Barcelona, Spain}
\author{Emilio Gallicchio}
\affiliation[NYC]{Department of Chemistry, Brooklyn College of the City University of New York;  PhD Program in Chemistry, Graduate Center of the City University of New York; PhD Program in Biochemistry, Graduate Center of the City University of New York, NY 11210, USA}
\author{John D. Chodera}
\affiliation[MSKCC]{Computational and Systems Biology Program, Sloan Kettering Institute, Memorial Sloan Kettering Cancer Center, New York, NY 10065, USA}
\author{Thomas E. Markland}
\affiliation[Stanford]{Department of Chemistry, Stanford University, 337 Campus Drive, Stanford, CA, 94305, USA}
\author{Gianni De Fabritiis}
\email{g.defabritiis@gmail.com}
\affiliation{Computational Science Laboratory, Universitat Pompeu Fabra, Barcelona Biomedical Research Park (PRBB), C Dr. Aiguader 88, 08003, Barcelona, Spain}
\alsoaffiliation{Acellera Labs, C Dr Trueta 183, 08005, Barcelona, Spain}
\alsoaffiliation{Instituci\'o Catalana de Recerca i Estudis Avan\c{c}ats (ICREA), Passeig Lluis Companys 23, 08010 Barcelona, Spain}

\begin{document}
\maketitle
\begin{abstract}
This letter gives results on improving protein-ligand binding affinity predictions based on molecular dynamics simulations using machine learning potentials with a hybrid neural network potential and molecular mechanics methodology (NNP/MM). We compute relative binding free energies (RBFE) with the Alchemical Transfer Method (ATM) and validate its performance against established benchmarks and find significant enhancements compared to conventional MM force fields like GAFF2.
\end{abstract}

\section{Introduction}
In modern drug discovery, alchemical free energy calculations have emerged as highly efficient tools. Relative binding free energy calculations are widely employed in hit-to-lead approaches, and several commercial and free tools with comparable performance have been developed over the years. However, the accuracy of binding free energy calculations is influenced by the choice of ligand force field. Most conventional force fields like GAFF\cite{wang2004development,wang2006automatic}, GenFF\cite{vanommeslaeghe2010charmm,vanommeslaeghe2012automation}, and OPLS\cite{roos2019opls3e} often rely on fixed charge molecular mechanics (MM). This lack of important energetic contributions limits their chemical accuracy and leads to poor modeling of torsions.\cite{ponder2003force,dauber2019biomolecular,hagler2019force}

To address these limitations, one approach involves using quantum mechanical (QM) levels of theory to model the ligands while treating the remaining environment with an MM force field in a hybrid potential.\cite{beierlein2011simple} However, QM/MM calculations are significantly more computationally expensive than MM calculations, posing challenges for drug discovery settings where RBFE calculations may be required for dozens or even hundreds of ligands.
Recently, neural network potentials (NNPs) have shown success in predicting QM energies with significantly reduced computational cost compared to QM methods. Notably, the ANI-2x\cite{ANI2x} model supports molecular systems comprising elements H, C, N, O, S, F, and Cl. Moreover, a hybrid method that integrates NNPs and MM, known as NNP/MM\cite{galvelis2023nnp}, has been developed, offering the potential to model ligands more accurately in RBFE calculations than traditional MM force fields.
The Alchemical Transfer Method (ATM) is a recently developed methodology for alchemical free energy calculations that we recently validated that allows an easy implementation of NNPs \cite{azimi2022relative}. In previous publications, this methodology with MM force fields on a robust dataset obtained similar results to other state-of-the-art methods such as FEP+.\cite{SabanesATM, ATMPsivant}.
In this work, we exploit the capabilities of ATM to test the hybrid approach of using ANI-2x\cite{ANI2x} as the neural network potential. Rufa et al. previously managed to reduce the error of absolute binding free energies from 0.97 to 0.47 kcal/mol for a congeneric ligand series for tyrosine kinase TYK2 by correcting the conventional MM simulation with an NNP/MM approach .\cite{Rufa2020.07.29.227959}ANI-2x has several limitations in terms of non-supporting charged molecules and certain elements but it is otherwise a useful test potential. Our main objective is to test the applicability of this methodology with different ligand force fields and to evaluate the feasibility of an NNP/MM approach in relative binding free energy calculations.

\section{Methods}
In this study, we evaluated a series of targets from both Wang's \textit{et al.}\cite{wang2015accurate} and Schindler's datasets.\cite{Merck_Benchmark} Due to the limitations of ANI-2x,\cite{ANI2x} the NNP of our choice in this study, there is a series of targets from the aforementioned datasets that cannot be computed due to the properties of its ligands. Consequently, we evaluated the following targets: Cyclin-dependent kinase 2 (CDK2), c-Jun N-terminal kinase 1 (JNK1), tyrosine kinase 2 (TYK2), P38 MAP kinase (P38), hypoxia-inducible transcription factor 2 (HIF2A), PFKFB3, spleen tyrosine kinase (SYK) and tankyrase 2 (TNKS2), totaling 301 ligand pairs. For the selected targets most of the ligands are compatible with ANI-2x, the rest (and its corresponding ligand pair calculations) were removed from the dataset. Due to the higher computational costs related to the integration of NNP into these calculations, a subset of all the possible ligand pairs to be evaluated was selected at random. 
The workflow in this project is similar to our previous work\cite{SabanesATM}. Protein and ligand structures were readily available from Wang's\cite{wang2015accurate} and Schindler's\cite{Merck_Benchmark} datasets. Ligands were parameterized with GAFF 2.11\cite{wang2004development,wang2006automatic}. The topologies were generated using the \textit{parameterize}\cite{galvelis2019scalable} tool. In contrast to our previous work, we now prepared complex systems using HTMD,\cite{doerr2016htmd} which automated and streamlined the preparation of multiple ligand pairs, along with the automatic selection of binding site residues. However, the manual selection of atom indexes for ligand alignment remained necessary.
The energy minimization, thermalization, and equilibration steps followed the procedures described in our previous work.\cite{SabanesATM} Additionally, the system was annealed to the symmetric alchemical intermediate ($\lambda$ = 1/2) for 250 ps. The classical RBFE simulations (GAFF2) were run in triplicate for each ligand pair running an ensemble of 60 ns per replica.
Concurrently, we performed the same calculations by using an NNP/MM approach.\cite{galvelis2023nnp} This hybrid method allowed us to simulate a portion of the molecular system (the small molecule) with an NNP, while the rest was simulated with MM, providing the ligands with optimized intra-molecular interactions. For both approaches, we used the Amber ff14SB parameters\cite{zou2019blinded,maier2015ff14sb} as well as the TIP3P water model. Classical RBFE simulations were run at a 4fs timestep while the NNP/MM runs were computed at 1fs timestep, both with the ATM integrator plugin\cite{sdm_plugin}. Hamiltonian replica exchange along the $\lambda$ space for each ATM leg was performed with the ASyncRE software\cite{gallicchio2015asynchronous}, specially customized for OpenMM and ATM.\cite{GitHubGallicchioLabAToMOpenMM}
Consistent with our previous work, we computed the binding free energies and their corresponding uncertainties from the perturbation energy samples using the Unbinned Weighted Histogram Analysis Method (UWHAM).\cite{tan2012theory} The resulting relative binding free energies ($\Delta \Delta G$) were compared to experimental measurements in terms of mean absolute error (MAE), root mean square error (RMSE), and Kendall Tau correlation coefficient. For all the possible systems, absolute $\Delta G$ values were computed with cinnabar, an analysis tool to compute absolute binding free energies from $\Delta\Delta G$ values via a maximum likelihood estimator.\cite{cinnabar} Cinnabar also generates the correlation plots and calculates the error and correlation statistics necessary. We compared the obtained values from calculations and the works by Wang \textit{et al}\cite{wang2015accurate} and Schindler \textit{et al}\cite{Merck_Benchmark} with FEP+.
To perform the calculations, we utilized the OpenMM-ML and NNPOps libraries on our in-house cluster comprising NVIDIA RTX 2080 Ti and NVIDIA RTX 4090 cards. Standard MM calculations were run on GPUGRID. The parallel replica exchange molecular dynamics simulations were conducted using the OpenMM 7.7 MD engine and the ATM Meta Force plugin, utilizing the CUDA platform.

\section{Results}
\begin{figure}[h!]
\centering
\includegraphics[width=\columnwidth]{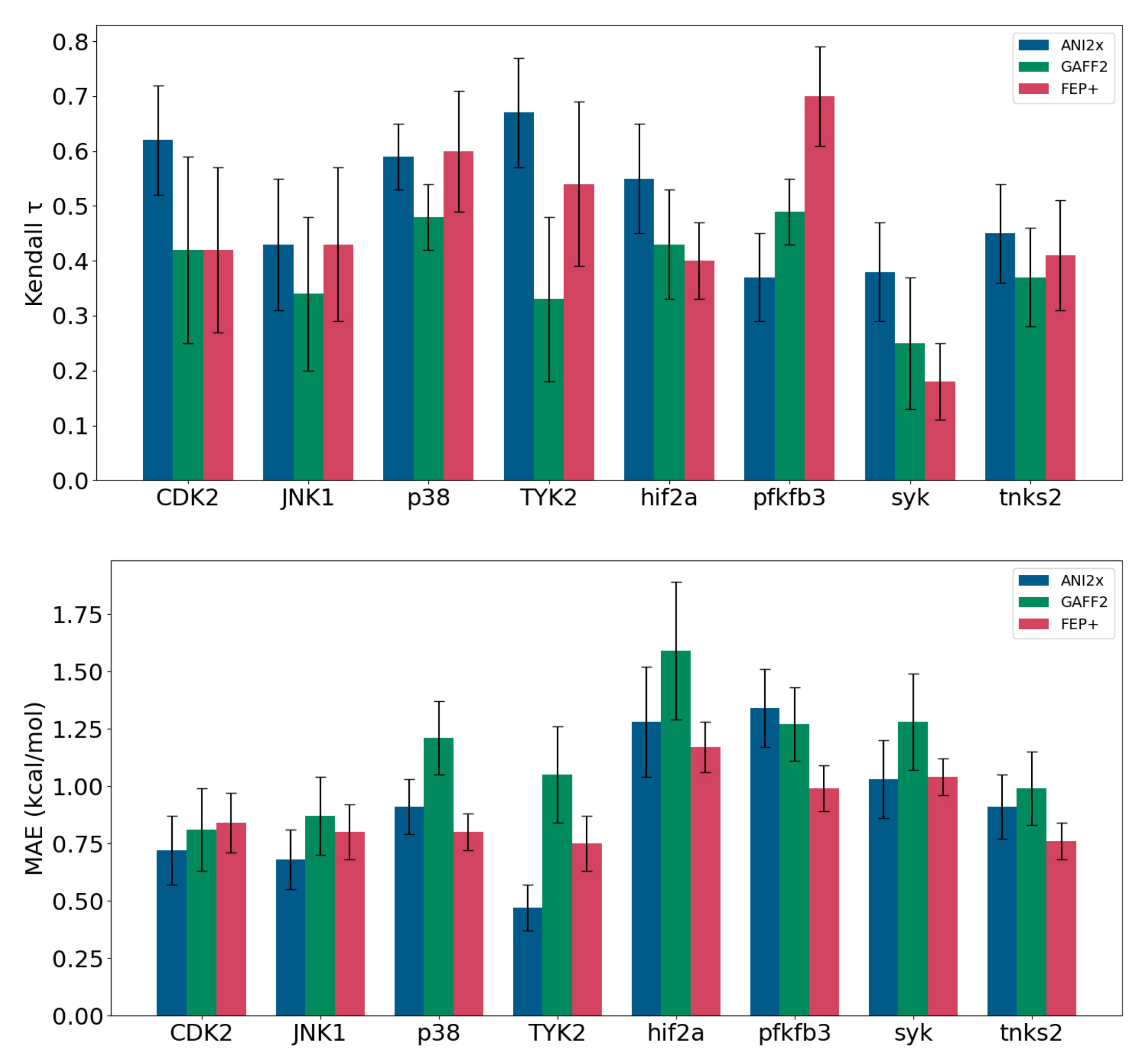}
\caption{ (Top) Kendall Tau and (bottom) Mean Absolute Error (MAE) for the $\Delta \Delta G$s of each protein-ligand system calculated in combination with different force fields and reported estimates using FEP+\cite{wang2015accurate}}
\label{fig:comparison_corrs}
\end{figure}

\begin{figure*}
\centering
\includegraphics[width=\linewidth]{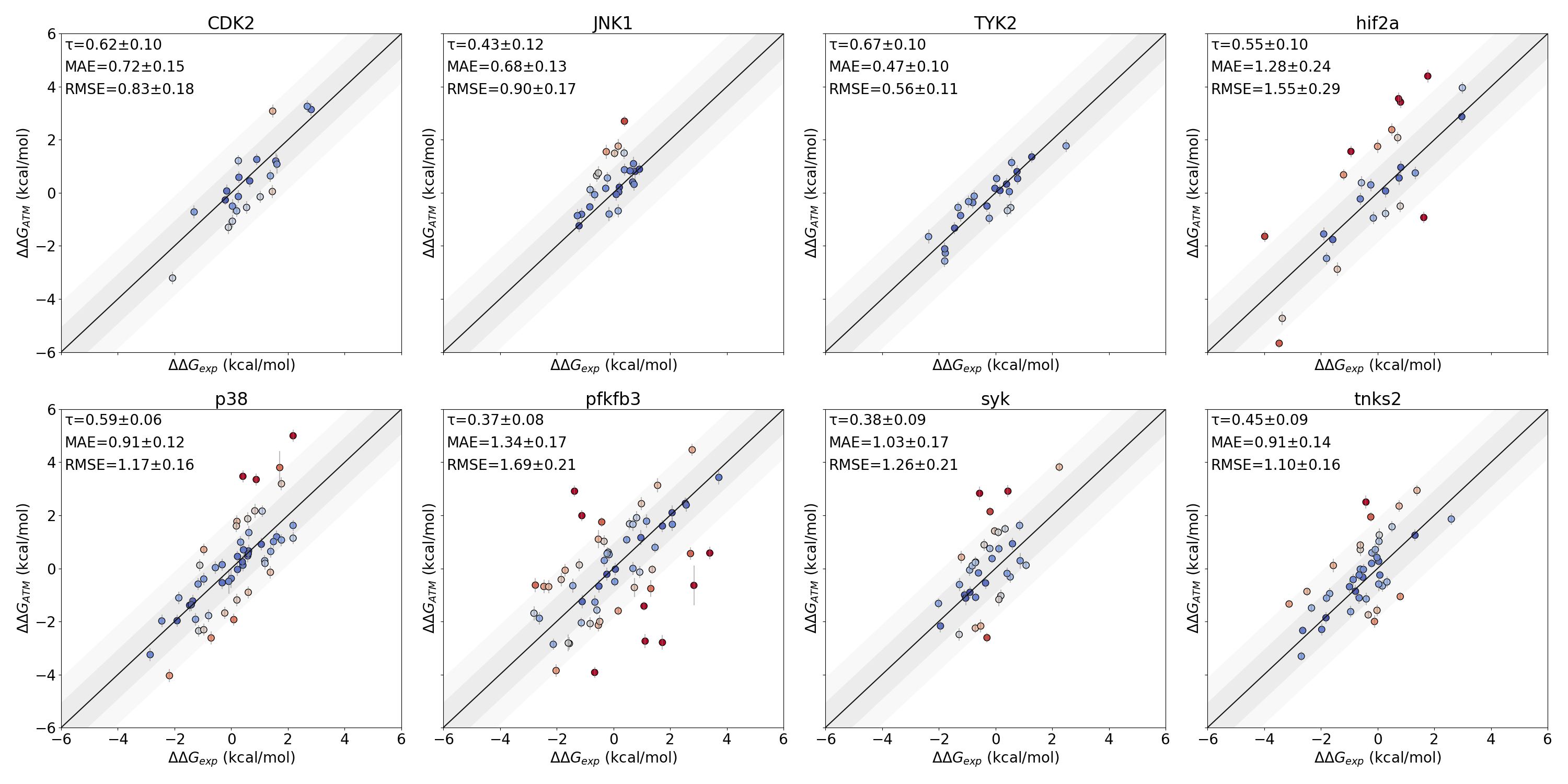}
\caption{Performance in combination with the neural network potential (NNP) for each protein-ligand system studied. The calculated $\Delta\Delta G$ estimates are plotted against their corresponding experimental values. MAE and RMSE are in kcal/mol and $\tau$ is Kendall correlation.}
\label{fig:prots_ATM}
\end{figure*}

The results of our simulations are displayed in Table \ref{tab:stats_table} and Figures \ref{fig:comparison_corrs} and \ref{fig:prots_ATM} which highlight the relative (Kendall's rank order correlation) and absolute performance (MAE and RMSE) of the evaluated methods. We do not report the Pearson correlation as well because the value is not significant for ($\Delta\Delta G$) values as it varies with the choice of the pairs\cite{hahn2022best}. We cannot calculate $\Delta G$  for all pairs since we ran a subset of the original datasets. The computation of $\Delta G$ for all ligands was not possible due to a poor connection of the perturbation network. Figures \ref{fig:DGs_compare}-\ref{fig:DGs_rho} displays the $\Delta G$ values and related statistics for the systems that were possible to compute. The NNP/MM method demonstrated superior performance over pure MM runs in both relative and absolute measures. We observe that NNP/MM shows a better correlation coefficient and MAE for all of the evaluated systems but PFKFB3 when compared to ATM with GAFF2 as a force field. In comparison to FEP+, NNP/MM has a lower correlation for two systems (P38 and PFKFB3) and higher MAE for four of them (P38, HIF2A, PFKFB3 and TNKS2). Furthermore, the amount of ligands that are more accurately predicted is increased. In comparison to our GAFF2 runs, MM/NNP predicts a higher percentage of ligands with a MAE lower than both 1 and 1.5 kcal/mol.(Table \ref{tab:MAE_pct_table}). Additionally, there are no major differences between the conformers generated with GAFF2 and ANI-2x as force-fields. (Figure \ref{fig:conformers_compare}) We observe for some specific cases how ligands that participated in poor predictions with GAFF2 (MAE $>$ 2kcal/mol) are now predicted correctly (MAE $<$ 1kcal/mol).(Table \ref{tab:TYK2_example}) However, this improvement comes at a cost, as NNP/MM calculations are slower than conventional MM calculations\cite{galvelis2023nnp}. For instance, an RTX 4090 could yield up to 27 ns/day, whereas an ATM conventional run for a P38 system with 49k atoms is able to compute 211 ns/day (Figure \ref{fig:ATM_speed}). This decrease on speed mainly arises due to the limitation of a 1fs timestep with the current ATM integrator. While there is a considerable increase in computational cost for NNP/MM runs, both approaches could benefit from further optimizations. We also evaluate if different timesteps could influence the accuracy of RBFE calculations. We compared the results of the GAFF2 calculations performed in this work with a 4fs timestep with the calculated points from our previous benchmark, that were run at a 2fs timestep.(Figure \ref{fig:fs_compare}) We do not observe any considerable accuracy difference between the calculations at both timesteps. In terms of convergence, we observed that 60 ns per calculation tends to be sufficient. Convergence analysis over time shows good convergence for most cases as illustrated in Figure \ref{fig:convergence_compare}.

\begin{table*}
\centering
\resizebox{\textwidth}{!}{%
\begin{tabular}{|l|cccc|cccc|cccc|c|}
\hline
& & & \textbf{GAFF2}& & & & \textbf{NNP/MM ANI2x}& & & & \textbf{FEP+}& & \textbf{ligand pairs} \\
 & \textbf{kendall ($\tau$)} & & \textbf{MAE} & \textbf{RMSE} & \textbf{kendall ($\tau$)} & & \textbf{MAE} & \textbf{RMSE} & \textbf{kendall ($\tau$)} & & \textbf{MAE} & \textbf{RMSE} & \\
\hline
CDK2 & 0.42 $\pm$ 0.17 & & 0.8 $\pm$ 0.2 & 1.1 $\pm$ 0.3 & \textbf{0.62 $\pm$ 0.10} & & \textbf{0.7 $\pm$ 0.1} & \textbf{0.8 $\pm$ 0.2} & 0.42 $\pm$ 0.15 & & 0.8 $\pm$ 0.1 & 1.1 $\pm$ 0.1 & 22 \\ 
JNK1 & 0.34 $\pm$ 0.14 & & 0.9 $\pm$ 0.2 & 1.0 $\pm$ 0.2 & \textbf{0.43 $\pm$ 0.12} & & \textbf{0.7 $\pm$ 0.1} & \textbf{0.9 $\pm$ 0.2} & 0.43 $\pm$ 0.14 & & 0.8 $\pm$ 0.1 & 1.0 $\pm$ 0.1 & 27 \\ 
P38 & 0.48 $\pm$ 0.06 & & 1.2 $\pm$ 0.2 & 1.6 $\pm$ 0.2 & 0.59 $\pm$ 0.05 & & 0.9 $\pm$ 0.1 & 1.2 $\pm$ 0.2 & \textbf{0.60 $\pm$ 0.11} &&\textbf{ 0.8 $\pm$ 0.1} & \textbf{1.0 $\pm$ 0.1} & 56 \\ 
TYK2 & 0.33 $\pm$ 0.15 & & 1.1 $\pm$ 0.2 & 1.3 $\pm$ 0.3 & \textbf{0.67 $\pm$ 0.10} & &\textbf{ 0.5 $\pm$ 0.1} &\textbf{ 0.6 $\pm$ 0.1} & 0.54 $\pm$ 0.15 & & 0.8 $\pm$ 0.2 & 0.9 $\pm$ 0.1 & 24 \\ 
HIF2A & 0.43 $\pm$ 0.10 & & 1.6 $\pm$ 0.3 & 2.0 $\pm$ 0.4 & \textbf{0.55 $\pm$ 0.11} && 1.3 $\pm$ 0.2 & 1.6 $\pm$ 0.3 & 0.50 $\pm$ 0.13 & & \textbf{1.1 $\pm$ 0.1} & \textbf{1.3 $\pm$ 0.2} & 28 \\ 
PFKFB3 & 0.49 $\pm$ 0.06 & & 1.3 $\pm$ 0.2 & 1.6 $\pm$ 0.2 & 0.37 $\pm$ 0.08 & & 1.3 $\pm$ 0.2 & 1.7 $\pm$ 0.2 & \textbf{0.70 $\pm$ 0.09} & & \textbf{1.0 $\pm$ 0.1 }& \textbf{1.6 $\pm$ 0.2} & 62 \\ 
SYK & 0.25 $\pm$ 0.12 & & 1.3 $\pm$ 0.2 & 1.6 $\pm$ 0.3 & \textbf{0.38 $\pm$ 0.10} & &\textbf{ 1.0 $\pm$ 0.2} & \textbf{1.3 $\pm$ 0.2} & 0.16 $\pm$ 0.11 & & 1.2 $\pm$ 0.1 & 1.5 $\pm$ 0.2 & 37 \\ 
TNKS2 & 0.37 $\pm$ 0.09 & & 1.0 $\pm$ 0.2 & 1.2 $\pm$ 0.2 & \textbf{0.45 $\pm$ 0.10} && 0.9 $\pm$ 0.1 & \textbf{1.1 $\pm$ 0.2} & 0.41 $\pm$ 0.10 & & \textbf{0.8 $\pm$ 0.1} & 1.0 $\pm$ 0.1 & 45 \\ 
\hline
\end{tabular}%
}
\caption{\label{tab:stats_table} Comparison of the performance of different forcefields and NNP/MM. Kendall correlation ($\tau$), Mean Absolute Error (MAE) and Root Mean Square Error (RMSE) in kcal/mol for the 8 tested Protein Targets. FEP+ is included as a state-of-the-art comparison. }
\end{table*}

\section{Conclusion}
We conducted relative binding free energy (RBFE) calculations using an innovative NNP/MM approach. 
Our findings demonstrate the substantial accuracy enhancement achieved by using an NNP/MM approach at the cost of increased computational time. Compared to conventional ligand forcefields like GAFF2, the NNP/MM approach exhibited reduced mean absolute errors, with most systems reaching below 1 kcal/mol.
However, we acknowledge that the current NNP used in this study is limited to neutral molecules and a limited set of elements, posing a constraint on our exploration of the vast chemical space. Future endeavors should focus on expanding the applicability of NNPs to include charged ligands, thereby broadening the scope of our investigations. An increase in computing performance is also needed, probably with the inclusion of other integrators that allow for higher timesteps. Due to limited computational resources a random subset of all the possible calculations was computed. Although a potential bias could be included due to the nature of the subset we believe to have a sampled an extensive number of data points to understand the capabilities of RBFE along with the NNP/MM approach.
Our work highlights the potential of NNP/MM for accurate RBFE calculations and underscores the importance of further advancing NNPs to encompass a broader range of molecular species, and further improve the accuracy of these calculations.
\section{Data and software availability}
The calculated free energy values, ligand and protein structures, as well as preparation scripts, are available at \url{https://github.com/compsciencelab/ATM_benchmark/tree/main/ATM_With_NNPs}
\section{Acknowledgement}
The authors thank the volunteers of GPUGRID.net for donating computing time. This project has received funding from
the European Union’s Horizon 2020 research and innovation programme under grant agreement No. 823712;
and the project PID2020-116564GB-I00 has been funded by MCIN / AEI / 10.13039/501100011033; the Torres-Quevedo Programme from the Spanish National Agency for Research (PTQ2020-011145 / AEI / 10.13039/501100011033).
Research reported in this publication was supported by the National Institute of General Medical Sciences (NIGMS) of the National Institutes of Health under award number R01GM140090. 
{E. G.} acknowledges support from the United States' National Science Foundation (NSF CAREER 1750511).
J.D.C.\ acknowledges support from NIH grant P30CA008748, R01GM140090, and the Sloan Kettering Institute.
The content is solely the responsibility of the authors and does not necessarily represent the official views of the National Institutes of Health.
\section{Associated Content}
The Supporting Information contains:
Description of the workflow used for this work. Barplots of the pearson correlation and RMSE from the calculated versus experimental $\Delta \Delta G$ values for all systems and compared methods. Scatterplots for the calculated $\Delta G$ on all the connected systems and the comparison between the compared methods as well as barplots from the corresponding MAE, RMSE errors and R2 and spearman correlations. Barplots with the analysis of the different speed performances between ATM and ATM combined with NNP/MM method. Scatterplots for a series of targets studied at different timesteps. Convergence analaysis based on the estimation of the $\Delta \Delta G$ through simulated time.
\section{Disclosures}
J.D.C.\ is a current member of the Scientific Advisory Board of OpenEye Scientific Software, Redesign Science, Ventus Therapeutics, and Interline Therapeutics, and has equity interests in Redesign Science and Interline Therapeutics. 
The Chodera laboratory receives or has received funding from multiple sources, including the National Institutes of Health, the National Science Foundation, the Parker Institute for Cancer Immunotherapy, Relay Therapeutics, Entasis Therapeutics, Silicon Therapeutics, EMD Serono (Merck KGaA), AstraZeneca, Vir Biotechnology, Bayer, XtalPi, Interline Therapeutics, the Molecular Sciences Software Institute, the Starr Cancer Consortium, the Open Force Field Consortium, Cycle for Survival, a Louis V. Gerstner Young Investigator Award, and the Sloan Kettering Institute. A complete funding history for the Chodera lab can be found at \url{http://choderalab.org/funding}.

\clearpage
\onecolumn

\bibliography{ref}


\section{Supporting Information} \label{sec:sup_methods}
\setcounter{figure}{0}  
\setcounter{table}{0}  
\renewcommand{\thefigure}{S\arabic{figure}}
\renewcommand{\thetable}{S\arabic{table}}  

\subsection*{Supporting Methods}
\begin{figure}
\centering
\includegraphics[width=\columnwidth]{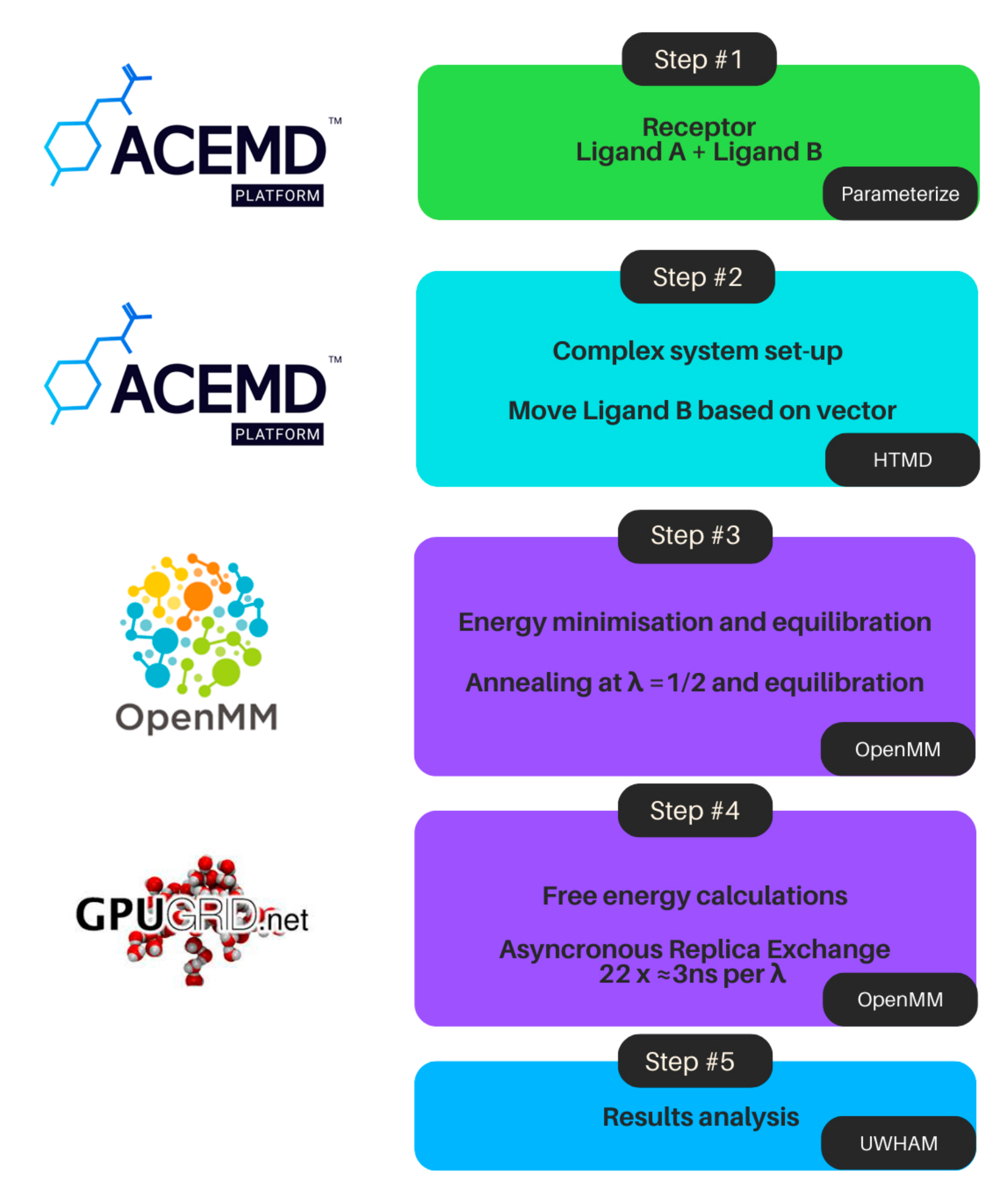}
\caption{The ATM workflow used in this work. Ligands topologies are calculated with \textit{parameterize} with GAFF2 and Sage force fields. (2) System complexes are prepared and built with htmd\cite{doerr2016htmd}. Protein topologies are prepared with the Amber ff14SB force field. Next ligand B is displaced based on a vector. (3) Energy minimization and equilibration is performed. Later an annealing and equilibration at $\lambda$=1/2 is performed. (4) Replica Exchange simulations are performed for a total sampling of 60ns. ATM simulations were run in GPUGRID were as ATM-NNP calculations were performed in our local cluster.(5) After the simulations were finished, these were analyzed with the UWHAM package to obtain the calculated $\Delta\Delta G$ estimates.}
\label{fig:Workflow_ATM}
\end{figure}

\begin{figure*}
\centering
\includegraphics[width=\linewidth]{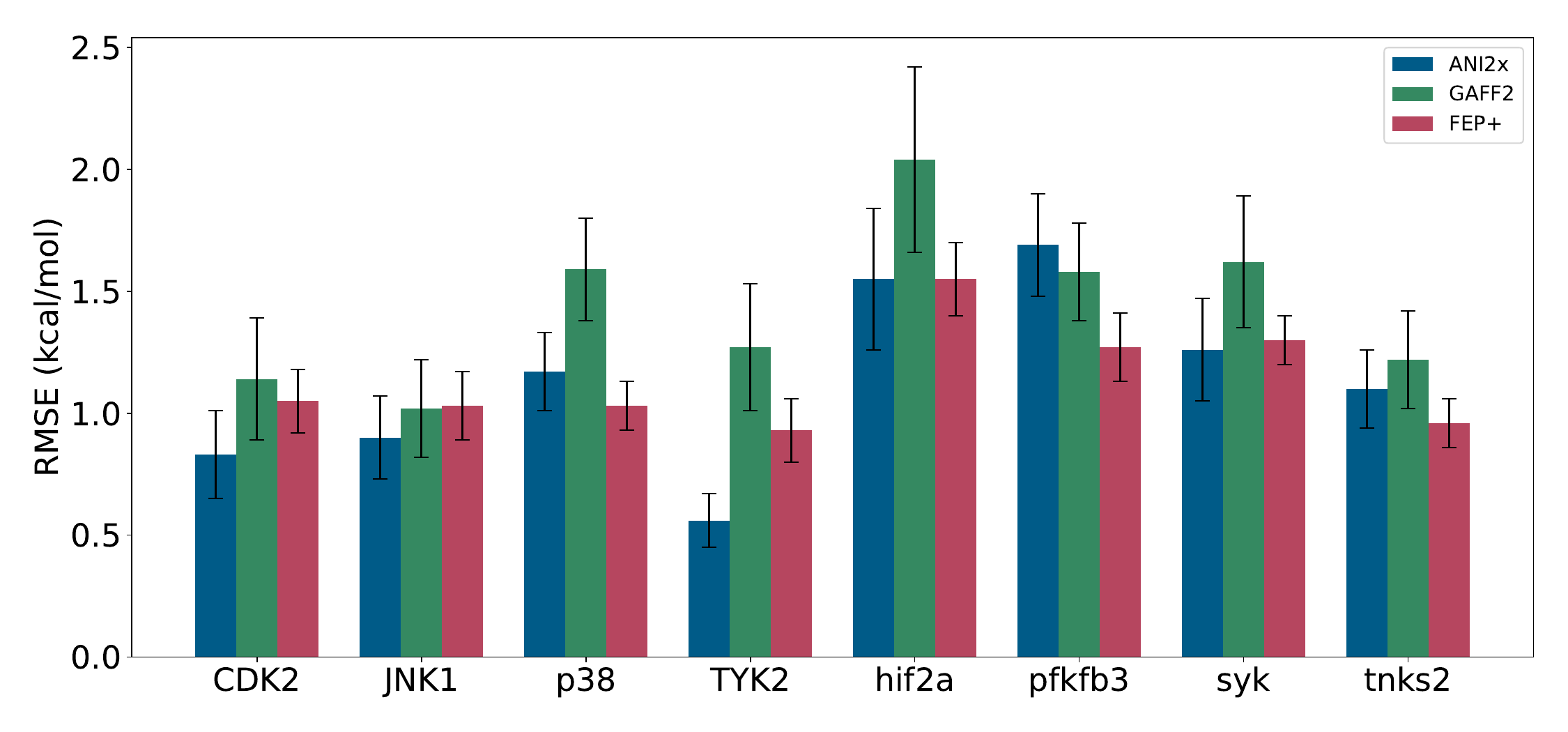}
\caption{Pearson correlation for each protein-ligand system calculated in combination with different force fields and reported estimates using FEP+}
\label{fig:r_ANI}
\end{figure*}

\begin{figure*}
\centering
\includegraphics[width=\linewidth]{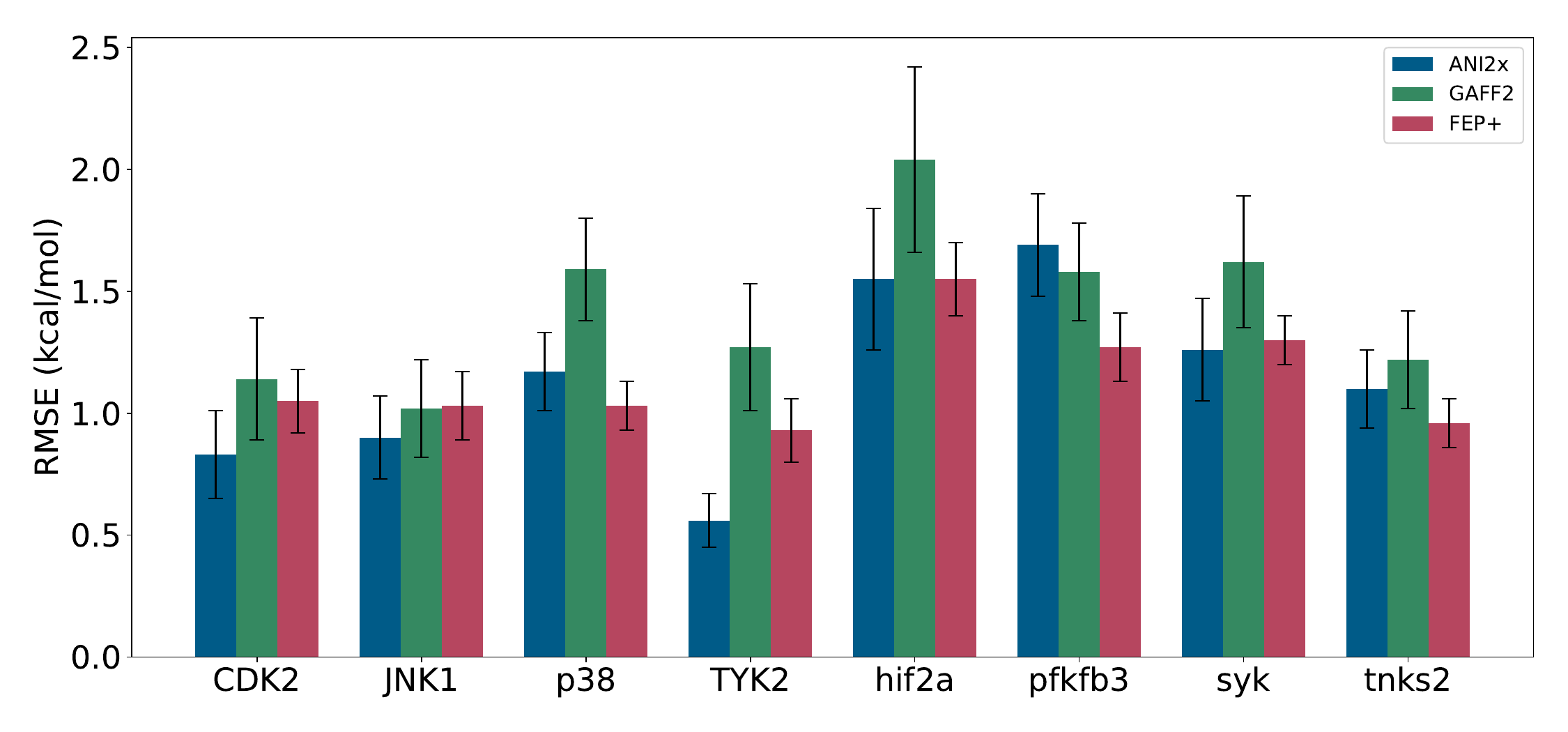}
\caption{Root Mean Square Error (RMSE) in kcal/mol for each protein-ligand system calculated in combination with different force fields and reported estimates using FEP+}
\label{fig:RMSE_ANI}
\end{figure*}

\begin{figure*}
\centering
\includegraphics[scale=0.38]{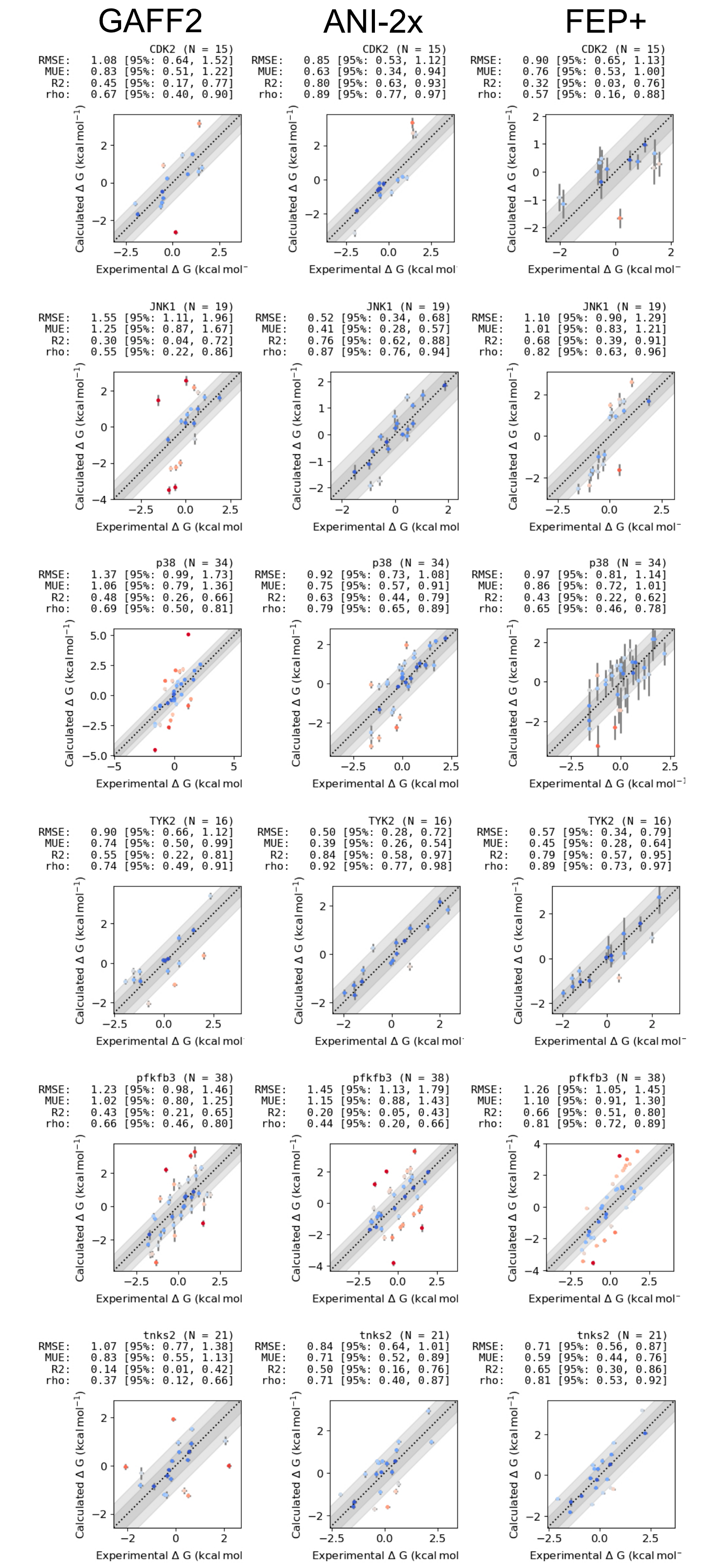}
\caption{Scatterplots for the $\Delta$G calculated on all the connected systems. Comparison between GAFF2, NNP/MM and FEP+. On top of each plot are the corresponding statistics.}
\label{fig:DGs_compare}
\end{figure*}

\begin{figure*}
\centering
\includegraphics[width=\linewidth]{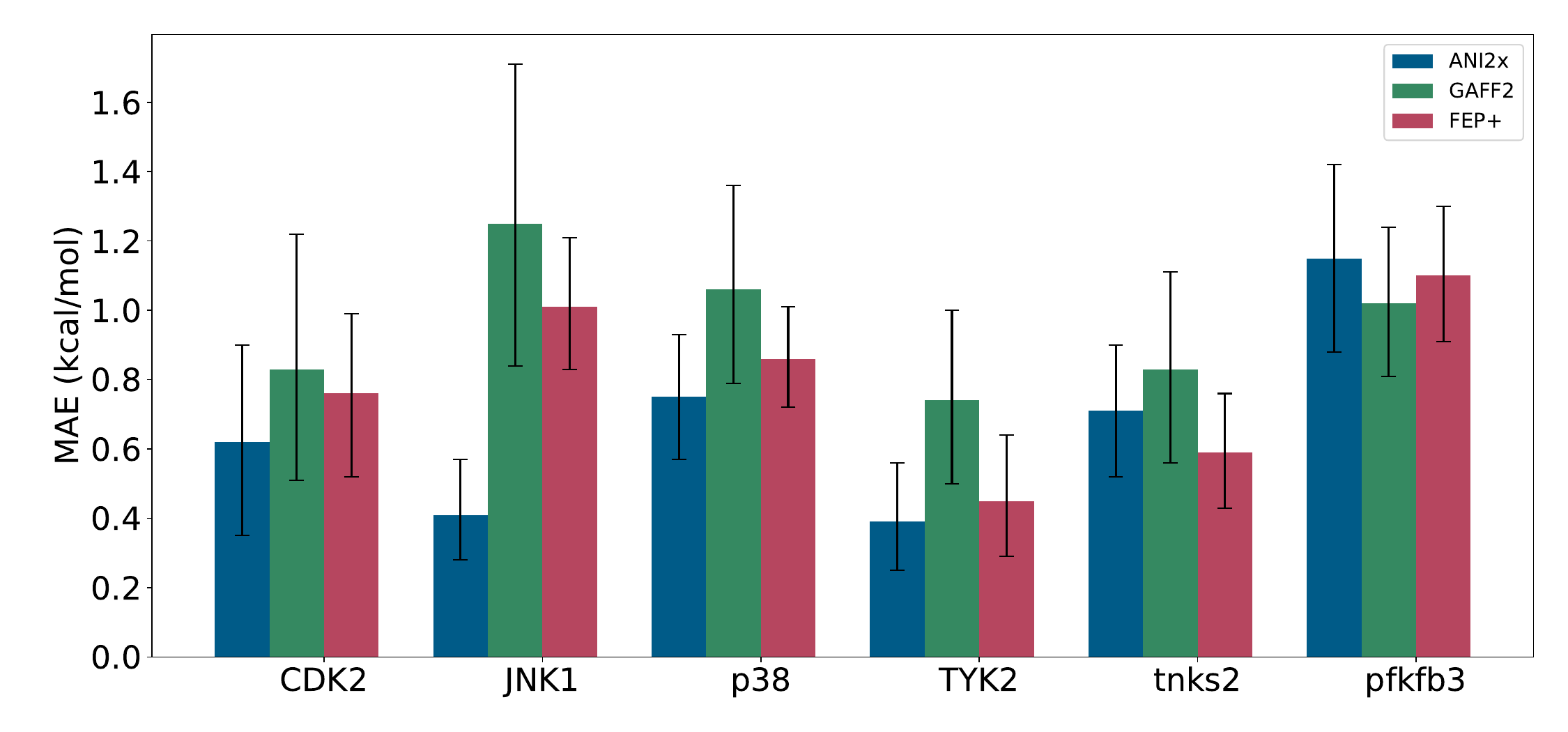}
\caption{MAE (kcal/mol) for the $\Delta$G values on all the connected systems}
\label{fig:DGs_MAE}
\end{figure*}

\begin{figure*}
\centering
\includegraphics[width=\linewidth]{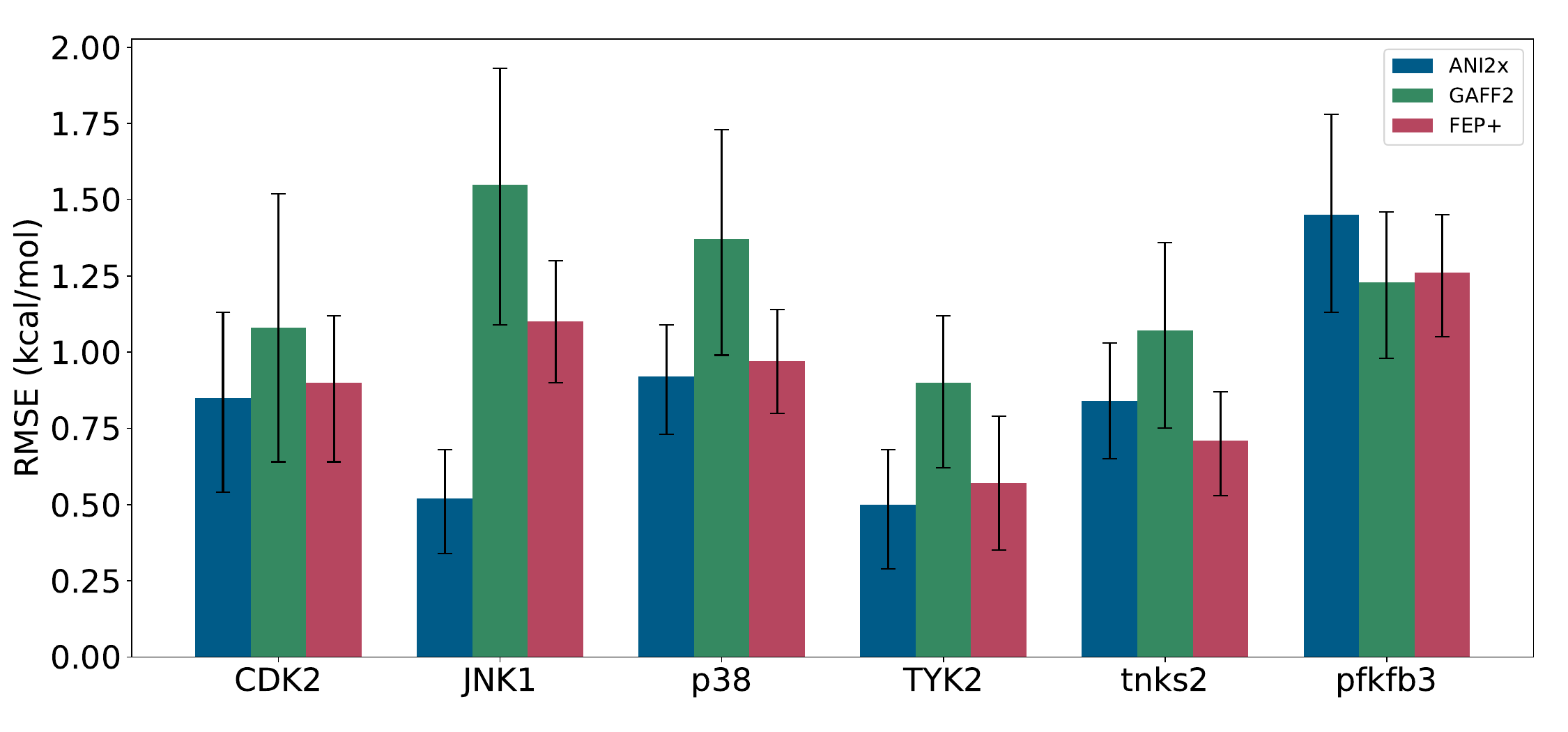}
\caption{RMSE (kcal/mol) for the $\Delta$G values on all the connected systems}
\label{fig:DGs_RMSE}
\end{figure*}

\begin{figure*}
\centering
\includegraphics[width=\linewidth]{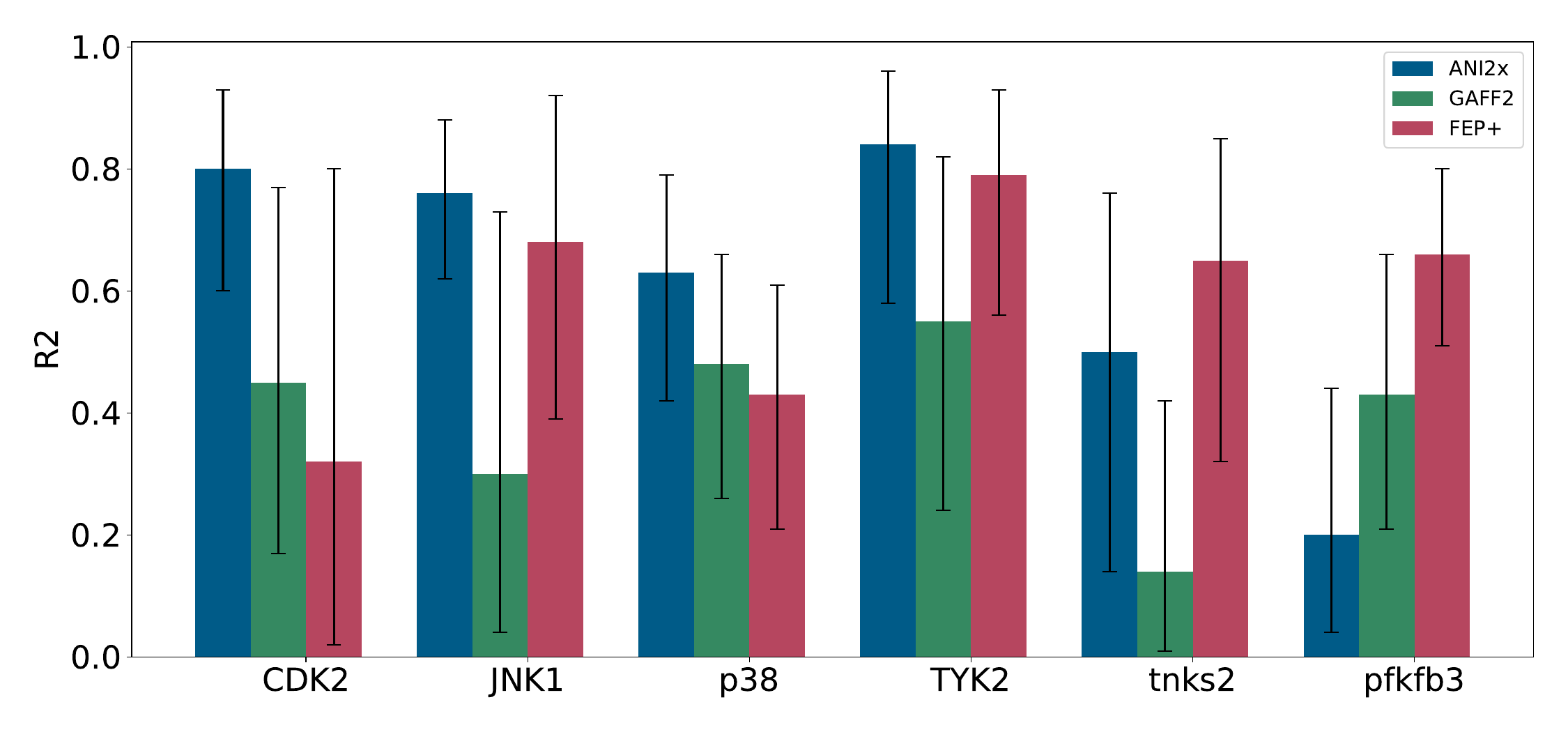}
\caption{R2 correlation for the $\Delta$G values on all the connected systems}
\label{fig:DGs_R2}
\end{figure*}

\begin{figure*}
\centering
\includegraphics[width=\linewidth]{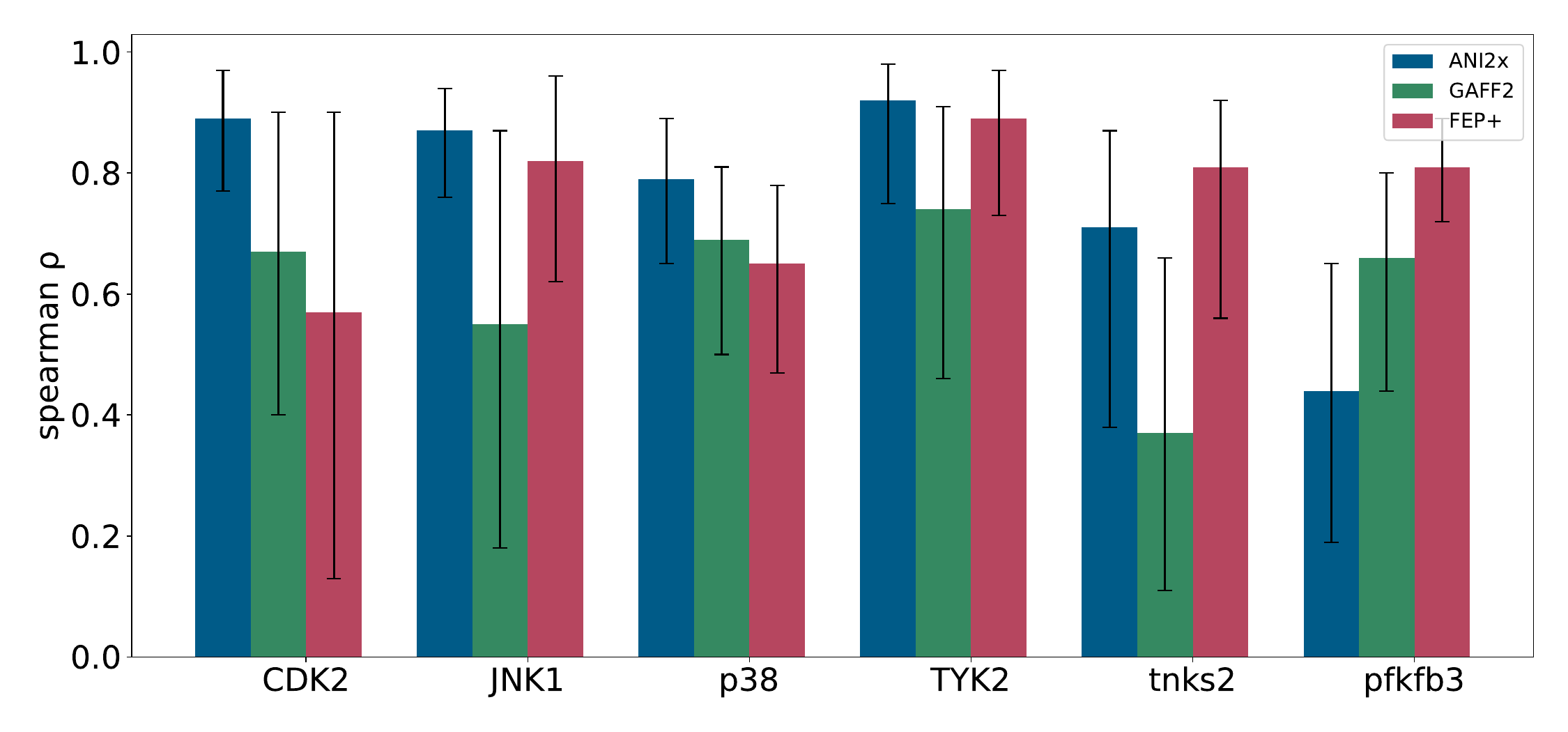}
\caption{Spearman correlation for the $\Delta$G values on all the connected systems}
\label{fig:DGs_rho}
\end{figure*}

\begin{figure*}
\centering
\includegraphics[width=\linewidth]{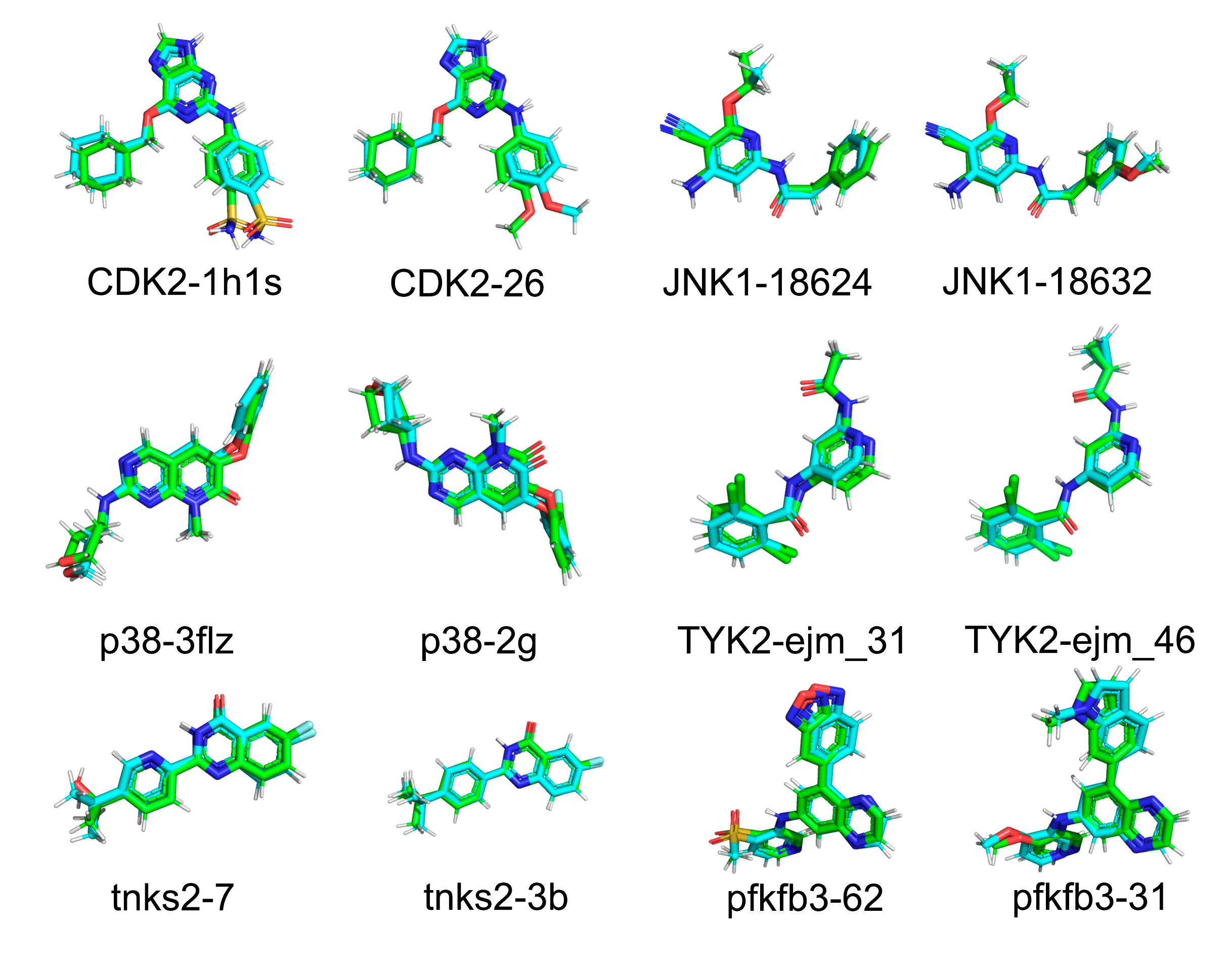}
\caption{Generated conformers after equilibration for runs performed with GAFF2 (cyan) and ANI-2x (green).}
\label{fig:conformers_compare}
\end{figure*}

\begin{figure*}
\centering
\includegraphics[width=\linewidth]{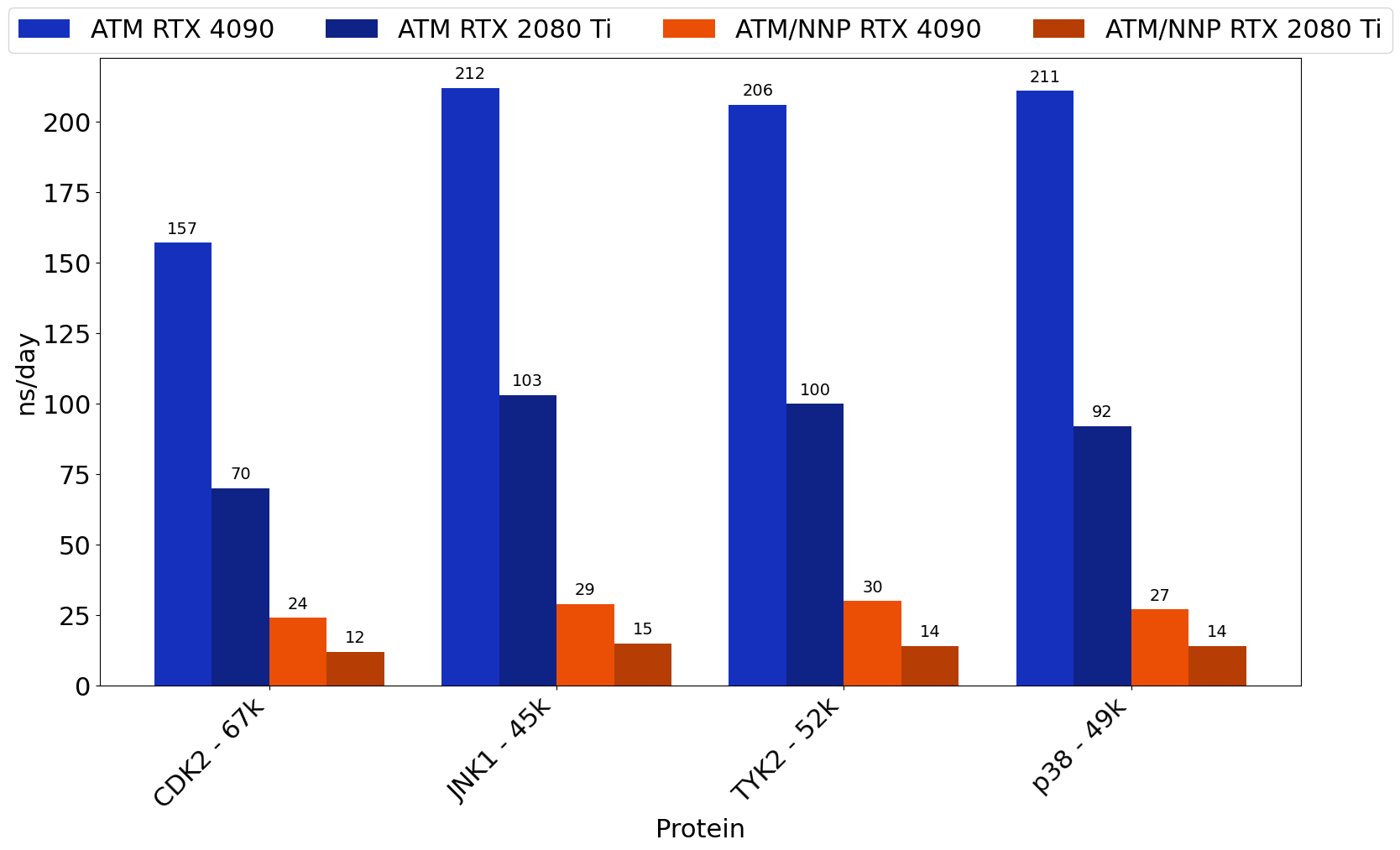}
\caption{Performance of ATM and ATM/NNP on RTX 2080Ti and RTX 4090 graphics cards with OpenMM 7.7 MD engine and the ATM Meta Force plugin using the CUDA platform}
\label{fig:ATM_speed}
\end{figure*}

\begin{figure*}
\centering
\includegraphics[width=\linewidth]{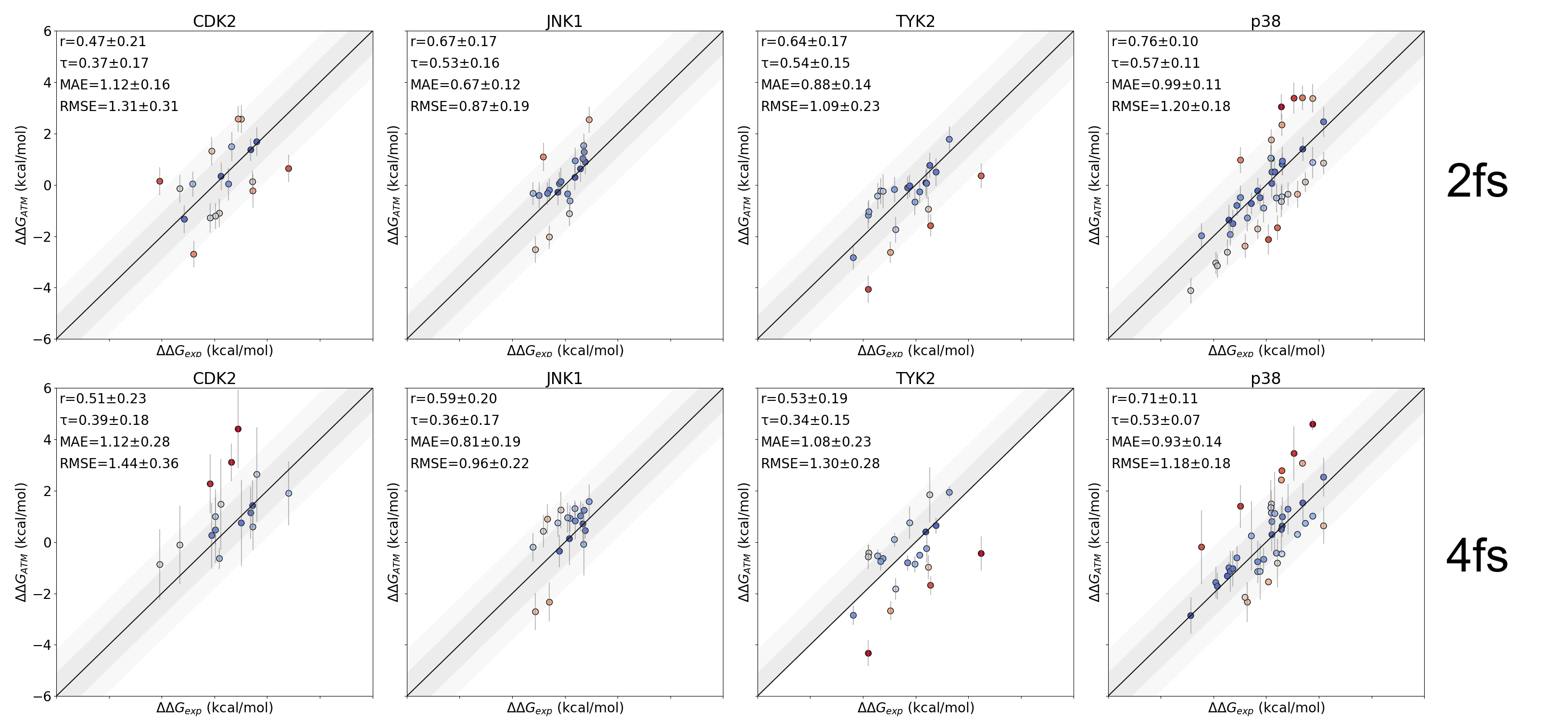}
\caption{Scatterplots for a series of targets studied at different timesteps. Top row are the relevant ligand pairs studied in our previous work, which we realized with a 2fs timestep. Bottom row are the calculations done for these targets at a 4fs timestep.}
\label{fig:fs_compare}
\end{figure*}

\begin{figure*}
\centering
\includegraphics[width=\linewidth]{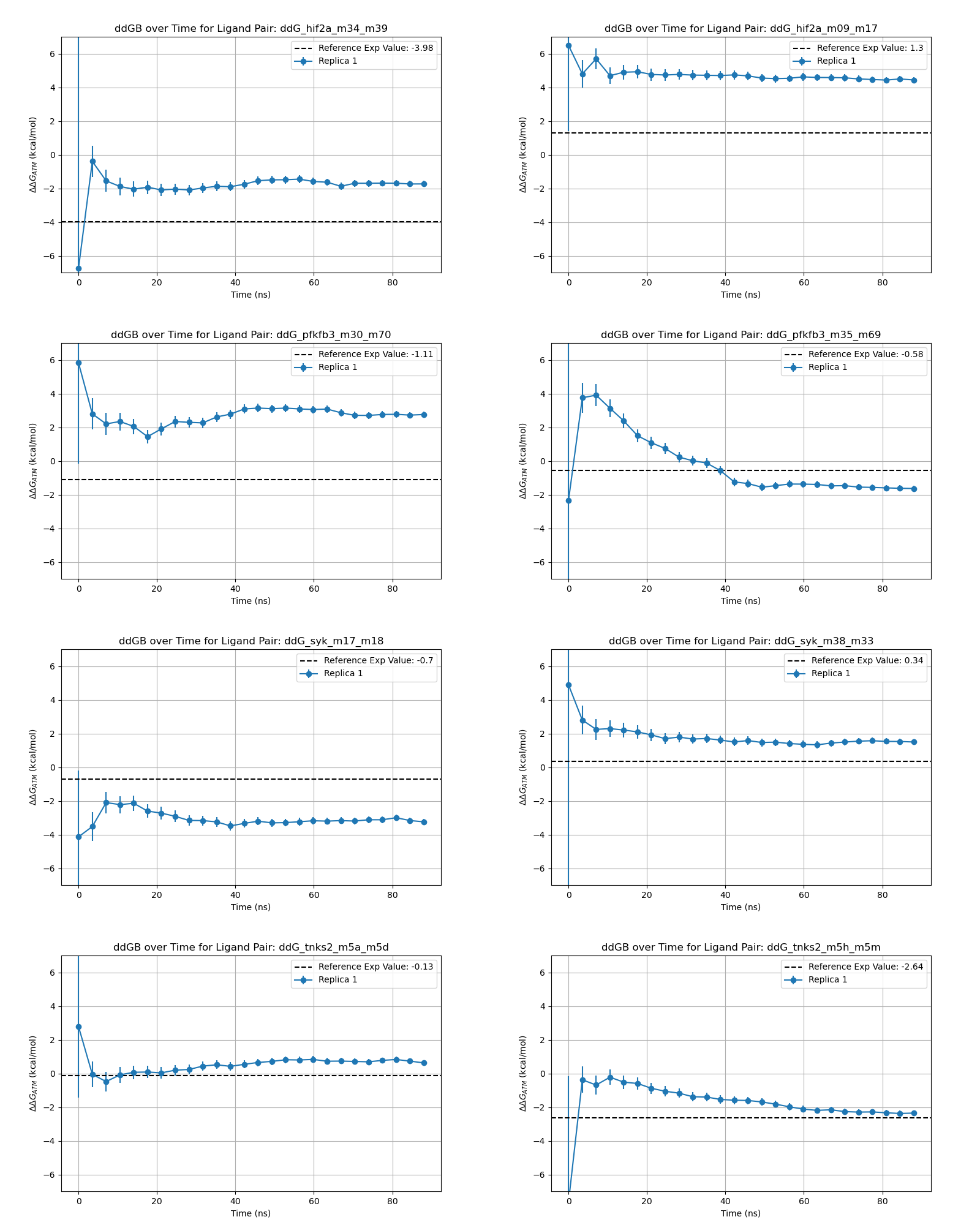}
\caption{Free energy convergence as a function of time for a series of ligand pairs of hif2a, pfkfb3, syk and tnks2}
\label{fig:convergence_compare}
\end{figure*}

\begin{figure*}
\centering
\includegraphics[width=\linewidth]{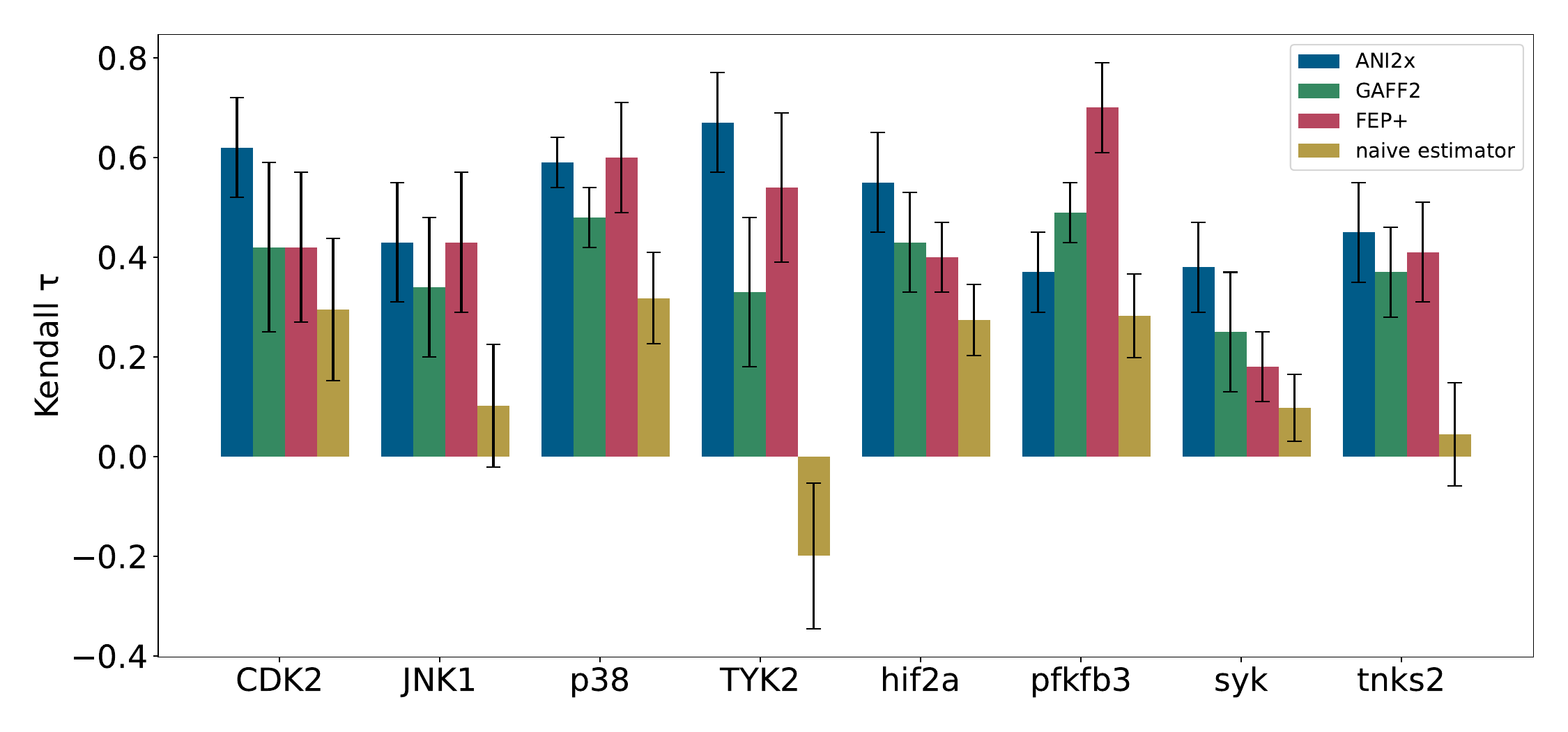}
\caption{Comparison of Kendall tau for the $\Delta \Delta G$s of each protein-ligand system calculated and compared against a naive estimator based on the difference of molecular weight between ligands}
\label{fig:naive_kendall}
\end{figure*}

\begin{table*}
\centering
\resizebox{\textwidth}{!}{%
\begin{tabular}{|p{1.5cm}|*{6}{p{2.8cm}|}}
\hline
{\textbf{Protein}} & \multicolumn{2}{c|}{\textbf{GAFF2}} & \multicolumn{2}{c|}{\textbf{NNP/MM ANI2x}} & \multicolumn{2}{c|}{\textbf{FEP+}} \\
 & \textbf{\% MAE $<$ 1} & \textbf{\% MAE $<$ 1.5} & \textbf{\% MAE $<$ 1} & \textbf{\% MAE $<$ 1.5} & \textbf{\% MAE $<$ 1} & \textbf{\% MAE $<$ 1.5} \\
\hline
\textbf{CDK2} & 54.6 $\pm$ 4.9 & 71.1 $\pm$ 4.5 & 64.0 $\pm$ 9.3 & 96.0 $\pm$ 3.9 & 54.5 $\pm$ 10.3 & 86.4 $\pm$ 7.2 \\
\textbf{JNK1} & 54.9 $\pm$ 4.7 & 79.6 $\pm$ 3.7 & 74.1 $\pm$ 8.3 & 88.9 $\pm$ 6.0 & 70.4 $\pm$ 8.9 & 85.2 $\pm$ 6.8 \\
\textbf{p38} & 49.9 $\pm$ 2.5 & 65.4 $\pm$ 2.4 & 55.9 $\pm$ 6.2 & 81.4 $\pm$ 5.0 & 64.3 $\pm$ 6.3 & 83.9 $\pm$ 4.8 \\
\textbf{TYK2} & 48.0 $\pm$ 3.6 & 68.4 $\pm$ 3.3 & 85.0 $\pm$ 5.6 & 97.5 $\pm$ 2.5 & 87.5 $\pm$ 7.9 & 87.5 $\pm$ 7.9 \\
\textbf{hif2a} & 36.0 $\pm$ 4.6 & 55.9 $\pm$ 4.9 & 41.9 $\pm$ 8.6 & 58.1 $\pm$ 8.8 & 55.2 $\pm$ 8.9 & 75.9 $\pm$ 8.1 \\
\textbf{pfkfb3} & 41.8 $\pm$ 3.6 & 60.4 $\pm$ 3.6 & 42.9 $\pm$ 6.2 & 65.1 $\pm$ 6.2 & 61.3 $\pm$ 6.1 & 80.6 $\pm$ 5.1 \\
\textbf{syk} & 40.4 $\pm$ 4.6 & 61.4 $\pm$ 4.6 & 59.5 $\pm$ 8.0 & 78.4 $\pm$ 6.8 & 42.1 $\pm$ 8.0 & 73.7 $\pm$ 7.1 \\
\textbf{tnks2} & 55.8 $\pm$ 5.2 & 70.5 $\pm$ 4.7 & 66.7 $\pm$ 7.1 & 77.8 $\pm$ 6.3 & 75.6 $\pm$ 6.2 & 88.9 $\pm$ 4.2 \\
\hline
\end{tabular}%
}
\caption{\label{tab:MAE_pct_table} Percentage of preditcions that have a MAE lower than 1 or 1.5 kcal/mol for each system. }
\end{table*}

\begin{table}
\centering

\begin{tabular}{|c|c|c|c|c |c |c |c |c|} \hline 
 \multicolumn{3}{|c|}{Protein: TYK2}& \multicolumn{3}{|c|}{NNP/MM ANI2x} & \multicolumn{3}{|c|}{GAFF2} \\ \hline 
ligand1 & ligand2 & exp\_ddG & ATM\_ddG & error & MAE & ATM\_ddG & error & MAE \\ \hline 
ejm\_31 & ejm\_46 & -1.77 & -2.27 & 0.25 & 0.50 & -0.42 & 0.24 & 1.35 \\ \hline 
ejm\_31 & ejm\_43 & 1.28 & 1.36 & 0.22 & 0.07 & 1.94 & 0.23 & 0.66 \\ \hline 
ejm\_31 & jmc\_28 & -1.44 & -1.33 & 0.22 & 0.11 & -0.54 & 0.23 & 0.90 \\ \hline 
ejm\_31 & ejm\_45 & -0.02 & 0.17 & 0.23 & 0.19 & -0.86 & 0.23 & 0.84 \\ \hline 
ejm\_31 & ejm\_48 & 0.54 & -0.56 & 0.24 & 1.10 & 1.84 & 0.24 & 1.30 \\ \hline 
ejm\_50 & ejm\_42 & -0.80 & -0.37 & 0.22 & 0.43 & 0.10 & 0.22 & 0.90 \\ \hline 
ejm\_55 & ejm\_54 & -1.32 & -0.55 & 0.22 & 0.77 & -0.76 & 0.23 & 0.56 \\ \hline 
ejm\_43 & ejm\_55 & -0.95 & -0.33 & 0.23 & \textbf{{\cellcolor{green!25}0.62}} & -2.68 & 0.23 & \textbf{{\cellcolor{red!25}1.73}} \\ \hline 
jmc\_28 & jmc\_30 & 0.04 & 0.57 & 0.26 & 0.53 & -1.07 & 0.28 & 1.11 \\ \hline 
jmc\_28 & jmc\_27 & -0.30 & -0.50 & 0.22 & 0.20 & -0.80 & 0.22 & 0.50 \\ \hline 
ejm\_49 & ejm\_31 & -1.79 & -2.57 & 0.24 & 0.78 & -0.57 & 0.24 & 1.22 \\ \hline 
ejm\_49 & ejm\_50 & -1.23 & -0.86 & 0.24 & 0.38 & -0.64 & 0.24 & 0.59 \\ \hline 
ejm\_45 & ejm\_42 & -0.22 & -0.96 & 0.22 & 0.74 & 0.75 & 0.23 & 0.97 \\ \hline 
ejm\_44 & ejm\_55 & -1.79 & -2.11 & 0.24 & \textbf{{\cellcolor{green!25}0.32}} & -4.33 & 0.23 & \textbf{{\cellcolor{red!25}2.54}} \\ \hline 
ejm\_44 & ejm\_42 & -2.36 & -1.65 & 0.27 & 0.71 & -2.85 & 0.24 & 0.49 \\ \hline 
ejm\_47 & ejm\_31 & 0.16 & 0.09 & 0.22 & 0.07 & -0.51 & 0.23 & 0.67 \\ \hline 
ejm\_47 & ejm\_55 & 0.49 & 0.04 & 0.22 & 0.44 & -0.98 & 0.23 & 1.47 \\ \hline 
jmc\_23 & jmc\_30 & 0.76 & 0.87 & 0.27 & 0.11 & -0.25 & 0.25 & 1.01 \\ \hline 
jmc\_23 & ejm\_46 & 0.39 & 0.33 & 0.22 & 0.06 & 0.40 & 0.22 & 0.01 \\ \hline 
jmc\_23 & ejm\_55 & 2.49 & 1.77 & 0.23 & \textbf{{\cellcolor{green!25}0.72}} & -0.44 & 0.23 & \textbf{{\cellcolor{red!25}2.93}} \\ \hline 
jmc\_23 & jmc\_27 & 0.42 & -0.67 & 0.24 & 1.09 & -0.25 & 0.22 & 0.67 \\ \hline 
ejm\_42 & ejm\_55 & 0.57 & 1.14 & 0.22 & \textbf{{\cellcolor{green!25}0.57}} & -1.68 & 0.22 & \textbf{{\cellcolor{red!25}2.25}} \\ \hline 
ejm\_42 & ejm\_48 & 0.78 & 0.53 & 0.22 & 0.25 & 0.64 & 0.23 & 0.14 \\ \hline 
ejm\_42 & ejm\_54 & -0.75 & -0.12 & 0.22 & 0.62 & -1.83 & 0.22 & 1.08 \\ \hline

\end{tabular}
\caption{\label{tab:TYK2_example} Case study example of the $\Delta\Delta$Gs obtained with NNP/MM and GAFF2. We observe how the transformations with the ligand ejm\_
55 give poor results with the GAFF2 (highlighted red) calculations but in the case of NNP/MM (highlighted green) the MAE is below 1kcal/mol.}

\end{table}

\end{document}